%% file: SPAWC_Invited_Paper.tex
\documentclass[conference]{IEEEtran}
\usepackage{amsmath,amssymb,dsfont,stfloats,color,url,mathtools}
\usepackage[pdftex]{graphicx}
\usepackage{wasysym,esint}
\usepackage{cite}
\usepackage{caption}
\usepackage{subcaption}
\usepackage{gensymb} 
\usepackage[boxruled,linesnumbered]{algorithm2e}
\usepackage{algorithm2e}
\usepackage{lipsum}

\DeclareGraphicsExtensions{.eps,.pdf,.png,.jpg,.gif,.jpeg,.pstex}
\newtheorem{rem}{Remark}

\input{macros}

\title{A Modular Pragmatic Architecture for Multiuser MIMO with Array-Fed RIS}

\author{\IEEEauthorblockN{Krishan K. Tiwari, Giuseppe Caire}
	\IEEEauthorblockA{Technische Universit\"at Berlin, 10587 Berlin, Germany}
	Email addresses: {lastname}@tu-berlin.de
}

\begin{document}

\bstctlcite{BSTcontrol} 

%\date{10 November 2021}
\maketitle

\begin{abstract}
We propose a power- and hardware-efficient, pragmatic, modular, multiuser/multibeam 
array-fed RIS architecture particularly suited to operate in very high frequency bands 
(high mmWave and sub-THz), where channels are typically sparse in the beamspace and line-of-sight (LOS) is required to achieve an acceptable received signal level. 
The key module is an active multi-antenna feeder (AMAF) with a small number of active antennas placed in the near field of a RIS with a much larger number of passive controllable reflecting elements. We propose a pragmatic approach to obtain a steerable beam with high gain and very low sidelobes. Then, $K$ independently controlled beams can be achieved by stacking $K$ of such AMAF-RIS modules. Our analysis takes in full account: 1) the near-end crosstalk (NEXT) between the modules, 2) the far-end crosstalk (FEXT) due to the sidelobes; 3) a thorough energy efficiency 
comparison with respect to conventional {\em active arrays} with the same beamforming performance.
Overall, we show that the proposed architecture is very attractive in terms of 
spectral efficiency, ease of implementation (hardware complexity), and energy efficiency.
\end{abstract} 

%While we show front illumination based RIS (Reflective Intelligent Surface) or reflectarray case, our proposed scheme/theory covers the back illumination based transmitarrays also since the back illumination has a reflected geometry with respect to the front illumination and under the idealized assumption made in our paper the two models are identical.

% The AMAF is much smaller than the legacy feed horn due to its higher aperture efficiency.

%%%%%%%%%%%%%%%%%%%%%%%%%%%%%%%%%%%%%%%%%%%%%%%%

\vspace{-0.5cm}
\section{Introduction}  
\label{sec:intro}
\vspace{-2pt}
%\cite{lens_cfg} to add ``our theory encompasses both since the back illumination has a reflected geometry with respect to the front illumination and 
%under the idealized assumption made in our paper the two models are identical. ''

Wireless communication in the millimeter wave (mmWave) and sub-THz frequency bands has garnered significant attention recently due to the promise of high data rates and ultra-low latency \cite{ted_thz}. At these frequencies, traditional wide-angle antennas and non-line-of-sight (NLOS) propagation are inadequate, prompting the use of large aperture antenna arrays for highly directional line-of-sight (LOS) propagation. This enables applications like wireless fronthaul, fixed point-to-multipoint wireless access (FWA), and LOS multiuser MIMO with highly directional beams \cite{JSDM}. However, the complexity and efficiency of large beamsteering active arrays remain problematic. To address this, innovative antenna configurations like Reflectarrays and Reflective Intelligent Surfaces (RIS) have been explored \cite{2018MRA, FAU_NF_RIS, ICC2023}. RIS, in particular, has been studied to modify wireless channels, but its effectiveness in the far field is limited by signal strength unless the RIS size is impractically large, even for indoor applications \cite{LozanoIndoor}.

\begin{figure}[h] 
\vspace{-0.3cm}
\centerline{\includegraphics[width=7.5cm]{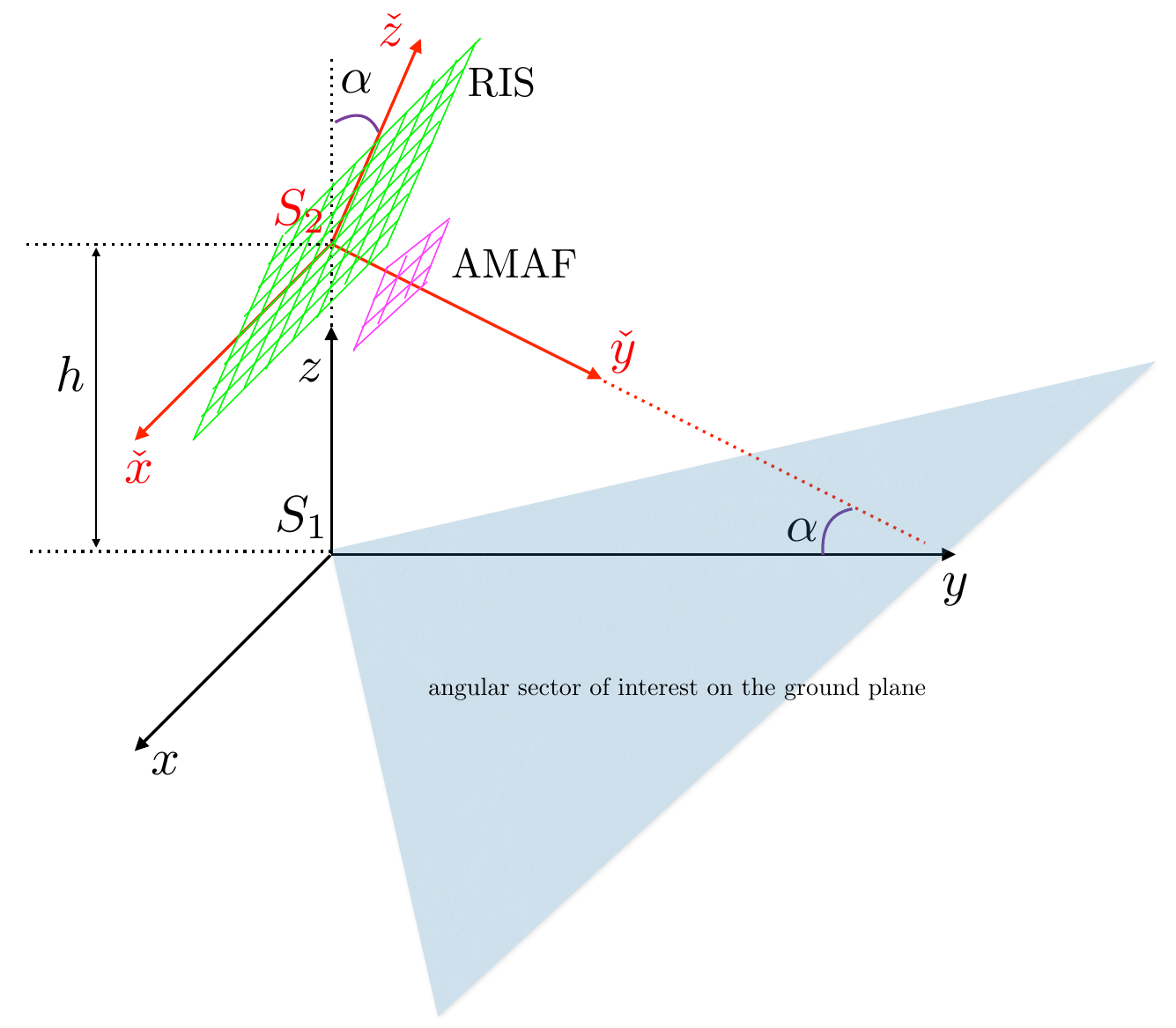}}
\caption{3D geometry, conceptual (not to scale).}
\label{3D-fig11}
\vspace{-0.3cm}
\end{figure}

From the Friis transmission equation \cite[eq. (1)]{Friis}, 
\vspace{-3pt}
\begin{equation} \label{fig:EEE}
\frac{P_{\rm rx}}{P_{\rm tx}} = \frac{A_{\rm tx} A_{\rm rx}}{(d \lambda)^2}, 
%\vspace{-0.5cm}
\end{equation}
where $\lambda$, $d$, $P$, and $A$ denote the carrier wavelength, distance, power, and antenna (effective) aperture, respectively and suffixes $_{\rm rx}$ and $_{\rm tx}$ denote receiver and transmitter. For fixed antenna size and decreasing $\lambda$, extreme power efficiency can be achieved. To achieve such gains at moderate hardware complexity, we propose the usage of RIS in a reflectarray configuration fed \cite{VTC2024} by a small active multiantenna feeder (AMAF).
%configured along the principal eigenmode of the AMAF-RIS propagation matrix.
In this paper, building on our previous 2D single module work \cite{ICC2023}, we present a full-dimensional (3D) multi-beam, multi-user model with planar RIS and AMAF arrays for terrestrial mmWave/sub-THz picocell base station applications, where the RIS is mechanically downtilted pointing to the picocell centroid on the ground, see Fig. \ref{3D-fig11}. 
Recall that the aperture efficiency of a planar patch array is much higher ($\approx 90\%$ due to a more uniform aperture distribution \cite[Fig. 6]{Munson74}) than that of ``legacy'' feed horn (51$\%$ for optimum pyramidal horns). In addition,  
the small size of the AMAF further alleviates the aperture blockage in the front illumination of reflectarrays. 
This motivates us to consider the AMAF instead of a horn as in traditional reflectarrays. Our mathematical model 
encompasses also the back illumination \cite{lens_cfg} which has a reflected geometry with respect to the front illumination. %since under the idealized assumption made in our work the two models are identical.
Whether front or back illumination is preferable depends on the blocking effect of the front feeder versus the power waste due to material absorption of the propagation through the metasurface \cite{lens_cfg}. These aspects are design/material dependent and go beyond the scope of this paper. 

\section{The AMAF-RIS module: design principles} 
\label{sec:3DModel}

As in Fig.~\ref{3D-fig11}, we define two coordinate systems. 
S1 has its origin on the ground plane x-y. 
S2 has its origin in the center of the RIS, positioned at $(0,0,h)$ with respect to the S1 system, and is downtilted by a rotation of $-\alpha$ in the z-y plane. Let $\iv = (1,0,0)^\transp,\jv = (0,1,0)^\transp, \kv = (0,0,1)^\transp$ denote the three versors of S1 in the coordinate system S1, and 
$\check{\iv} = (1,0,0)^\transp, \check{\jv} = (0,1,0)^\transp, \check{\kv} = (0,0,1)^\transp$ denote the three versors of S2 in the coordinate system S2. 
A S1 point $\pv = p_x \iv + p_y \jv + p_z \kv$ can be expressed in S2 Cartesian coordinates $\check{\pv} = \check{p}_x \check{\iv} + \check{p}_y \check{\jv} + \check{p}_z \check{\kv}$ by 
\begin{equation}
\check{\pv} = \left [ \begin{array}{ccc}
1 & 0 & 0 \\
0 & \cos(\alpha) & -\sin(\alpha) \\
0 & \sin(\alpha) & \cos(\alpha) \end{array} \right ] ( \pv - (0,0,h)^\transp).   \label{coord-transf}
\end{equation}

\begin{figure}[h!] 
\centerline{\includegraphics[width=8cm,height=4cm]{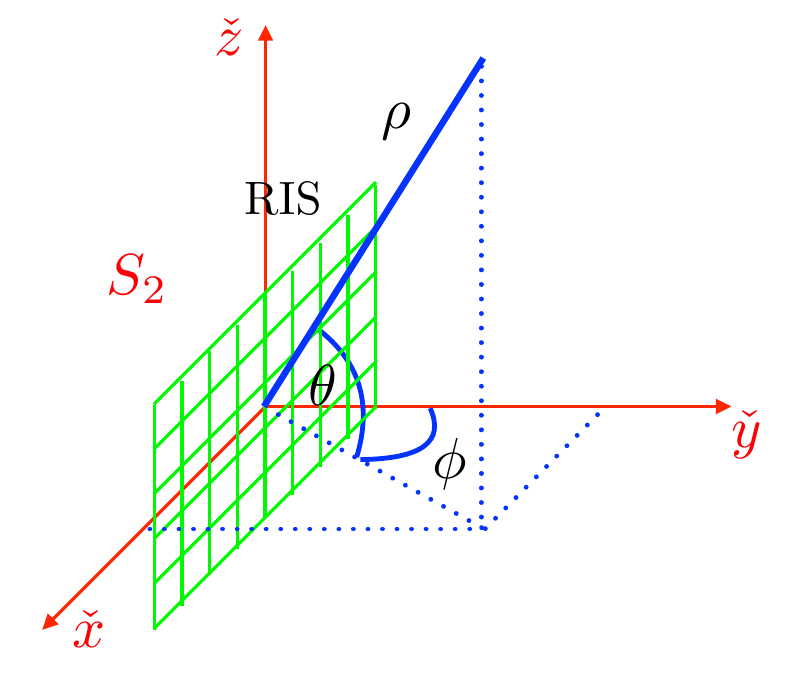}}
\caption{Spherical coordinates definition for the RIS coordinate system S2.}
\label{polarcoord1}
\end{figure}

For the S2 spherical coordinates in Fig.~\ref{polarcoord1}, the RIS boresight direction $\check{\jv}$ corresponds to the angles $\phi = 0, \theta = 0$, and this is a line tilted down and impinging the picocell centroid on the ground plane (S1 x-y plane) 
by an angle $\alpha$ (see Fig.~\ref{3D-fig11}). Hence, a point of $\check{\pv}$ in S2 with spherical coordinates $(\rho, \phi, \theta)$ has $\check{p}_x  = \rho \sin \phi \cos \theta,~ \check{p}_y  = \rho \cos \phi \cos \theta,~ \check{p}_z  =  \rho \sin \theta$. This, along with \eqref{coord-transf}, maps any S1 ground plane point to a far-field direction $(\phi,\theta)$ and a range $\rho$ with respect to the RIS. In the following, all lengths are normalized by $\lambda/2$. 

%%%%%%%%%%%%%%%%
%\vspace{-8pt}
\subsection{RIS Array and Far-Field Array Response}
\label{subsec:array}

A planar wavefront impinging a standard rectangular array (SRA) RIS at an angle $(\phi, \theta)$ has normal vector given by 
%\vspace{-5pt}
\begin{equation}
\nv(\phi, \theta) = (\sin \phi \cos \theta, \cos \phi \cos\theta,\sin\theta)^\transp.
\end{equation}
The planar wave propagation phase delay (complex phasor term) of the $(n,m)$ element of the RIS is given by 
%\vspace{-5pt}
    \begin{align}
    a_{n,m}(\phi,\theta)
    =  & \exp \left ( - j \pi \check{\pv}_{n,m}^\transp \nv(\phi,\theta) \right ). 
    \label{eq:array-vec}
    \end{align}
    where $\check{\pv}_{n,m}$ is the position of the $(n,m)$ RIS elements in the system S2. 
%\begin{equation} \label{eq:psi}
%\cos(\psi) = \nv(\phi,\theta) \cdot (0,1,0)^\transp = \cos \phi \cos \theta.
%\end{equation}
%$\left(\gamma(\in \mathbb{R}_+)<1 ~ \text{the AMAF-RIS power transfer coefficient}\right)$
We use axisymmetric model for the RIS and the AMAF patch elements (having a 3 dB beamwidth of $90 \degree$ and the power gain of 6 dBi) given by $G_{\rm patch}(\theta, \phi) = 4 \cos^2(\psi)$,   where $\psi$ is the angle of the direction $(\phi,\theta)$ with respect to the patch broadside, i.e., the S2 y-axis. Noticing that
$\cos(\psi) = \nv(\phi,\theta) \cdot (0,1,0)^\transp = \cos \phi \cos \theta$, we have
%\vspace{-10pt}
\begin{equation} \label{eq:patch}
G_{\rm patch}(\phi,\theta) = 4 \left ( \cos \phi \cos \theta \right )^2. 
\end{equation}
Assuming a RIS excitation by the AMAF such that the complex signal at each element is $u_{n,m}$, the RIS can further impose a  phase rotation $w_{n,m} = e^{j \mu_{n,m}}$ for each $n,m$ element. Hence, the resulting far-field radiation pattern as a function of 
the angle direction $(\phi,\theta)$ is given by 
%\vspace{-10pt}
\begin{equation} 
G(\phi,\theta) = G_{\rm patch}(\phi,\theta) \left | \sum_{n=0}^{N_x-1} \sum_{m = 0}^{N_z - 1} w_{n,m} u_{n,m} a_{n,m}^*(\phi,\theta) \right |^2.  \label{RISpattern}
\end{equation}
We also notice that, without loss of generality, we can incorporate the phase of 
$u_{n,m}$ into $w_{n,m}$. Hence, without loss of generality, we can 
replace $u_{n,m}$ by its magnitude $|u_{n,m}|$. 
For convenience, we define the ``tapered'' RIS weights 
$\tilde{w}_{n,m}(\phi,\theta) = |u_{n,m}| w_{n,m}$.
Collecting $\{a_{n,m}(\phi,\theta)\}$ and $\{\tilde{w}_{n,m}\}$ into two $N_p \times 1$ vectors $\av(\phi,\theta)$ and ${\tilde{\wv}}$, \eqref{RISpattern} can be 
compactly written as
%\vspace{-10pt}
\begin{equation} 
G(\phi,\theta) = 4 \left ( \cos \phi \cos \theta \right )^2 \left | \av(\phi,\theta)^\herm \tilde{\wv} \right |^2.\label{eq:bfpattern}
\end{equation}
%We define the RIS gain $\Gamma$ as the maximum possible value of $G(\phi,\theta)$ from (\ref{eq:bfpattern}). This is obtained
%for $\phi = \theta = 0$, letting $\wv = \onev$ (the all-one vector), and yields
%\begin{equation} 
%\Gamma = 4 \left | \sum_{n=0}^{N_x-1} \sum_{m = 0}^{N_z - 1} u_{n,m} \right |^2. \label{eq:gamma}
%\end{equation}

%In \cite{2013array}, it was shown that RIS radiation performance characterization using array theory closely matches the results from aperture theory, where full-wave simulations had shown that both methods can be used equivalently for a reliable calculation of the general pattern shape, main beam direction, beamwidth, and sidelobe levels. Therefore, we choose the array theory based RIS radiation pattern calculation as in (\ref{eq:bfpattern}), which is faster and requires less computational effort.

%%%%%%%%%%%%%%%%

\subsection{AMAF-RIS Illumination and Beam-Steering}
\label{subsec:T}

We focus now on the illumination, i.e., how to obtain a suitable (complex) 
amplitude profile $\{u_{n,m}\}$ induced by the AMAF on the RIS surface. The AMAF is formed by $N_a = N_h N_v$  active elements 
arranged in a $N_h \times N_v$ SRA and placed at a distance $F$ from the RIS with $N_p = N_x N_z$ elements arranged in a $N_x \times N_z$ SRA.
For convenience, we enumerate the RIS and the AMAF elements row by row using indices 
$k \in \{0, \ldots, N_p -1\}$ and $\ell \in \{0,\ldots, N_a - 1\}$, respectively. Letting $r_{k,\ell}$ denote the distance between 
the $k$-th RIS element and the $\ell$-th AMAF element, and letting $(\varphi_{k,\ell},\vartheta_{k,\ell})$ the angle at which they see each other with respect to their own normal (boresight) direction, narrowband AMAF-RIS propagation matrix $\Tm \in \CC^{N_p \times N_a}$ has (Friis formula and distance dependent phase term based) entries  
\begin{equation} %\label{eq:T}
T_{k,\ell}  = \frac{\sqrt{E_A(\varphi_{k,\ell},\vartheta_{k,\ell})E_R(\varphi_{k,\ell},\vartheta_{k,\ell}})}{2\pi r_{k,\ell}}~ e^{ - j\pi r_{k,\ell}},  \label{Telement}
\end{equation}
%\frac{\sqrt{E_A(\varphi_{k,\ell},\vartheta_{k,\ell})E_R(\varphi_{k,\ell},\vartheta_{k,\ell}})}{2\pi r_{k,\ell}}~ e^{ - j\pi r_{k,\ell}} e^{-j2\pi f \tau_{k,\ell}},
where $E_A(\varphi,\vartheta) = E_R(\varphi,\vartheta) = G_{\rm patch}(\varphi,\vartheta)$ in this work.

%$E_A(\varphi,\vartheta)$ and $E_R(\varphi,\vartheta)$.% are the AMAF and the RIS element radiation patterns, respectively. 
%In this work we assumed $E_A(\varphi,\vartheta) = E_R(\varphi,\vartheta) = G_{\rm patch}(\varphi,\vartheta)$. 

\begin{figure}[h] 
\centerline{\includegraphics[width=8cm,height=4cm]{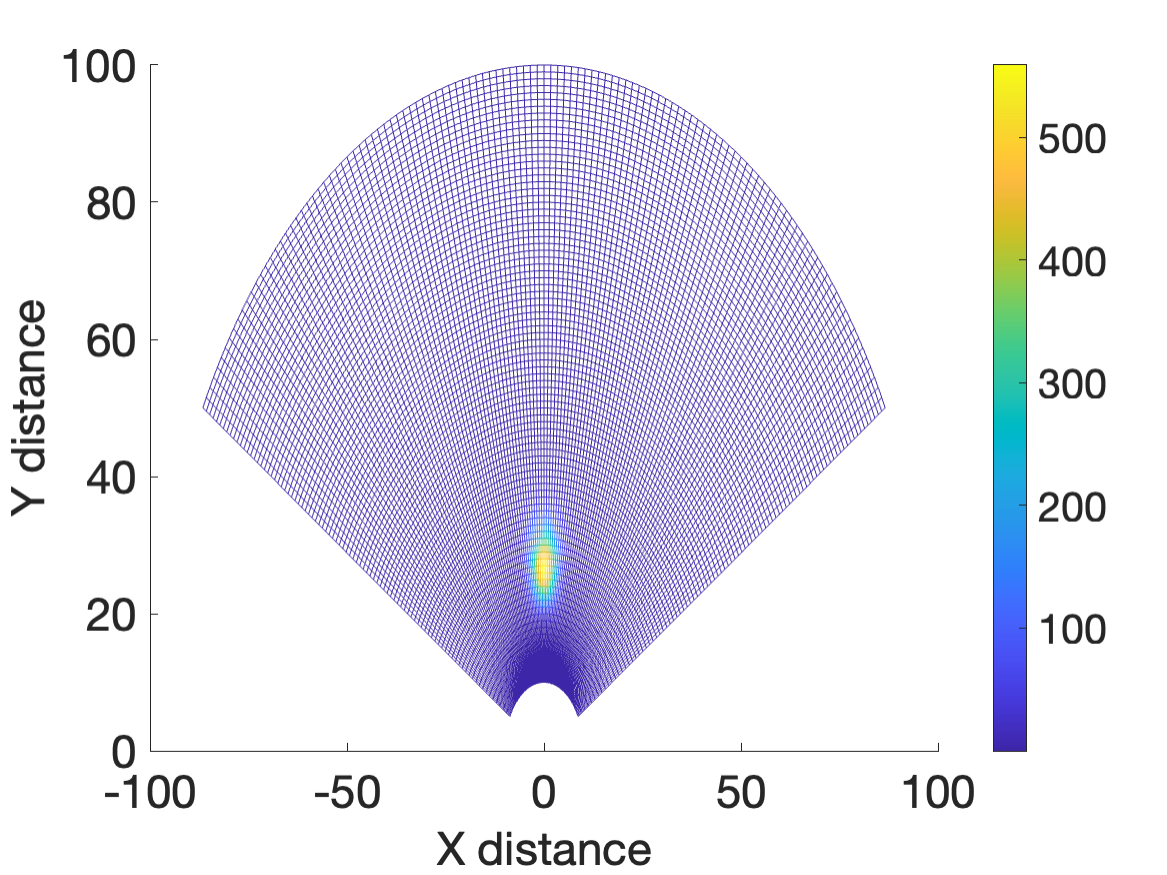}}
\caption{Ground footprint (coordinate system S1) of the RIS PEM spotbeam 
pointing at the RIS boresight (coordinate system S2), for a $16\times 16$ RIS with $2\times 2$ AMAF at distance $F = 6$.}
\label{fig:CenterSpot0}
\end{figure}

\begin{figure}[h] 
\vspace{-1cm}
\centerline{\includegraphics[width=6.5cm]{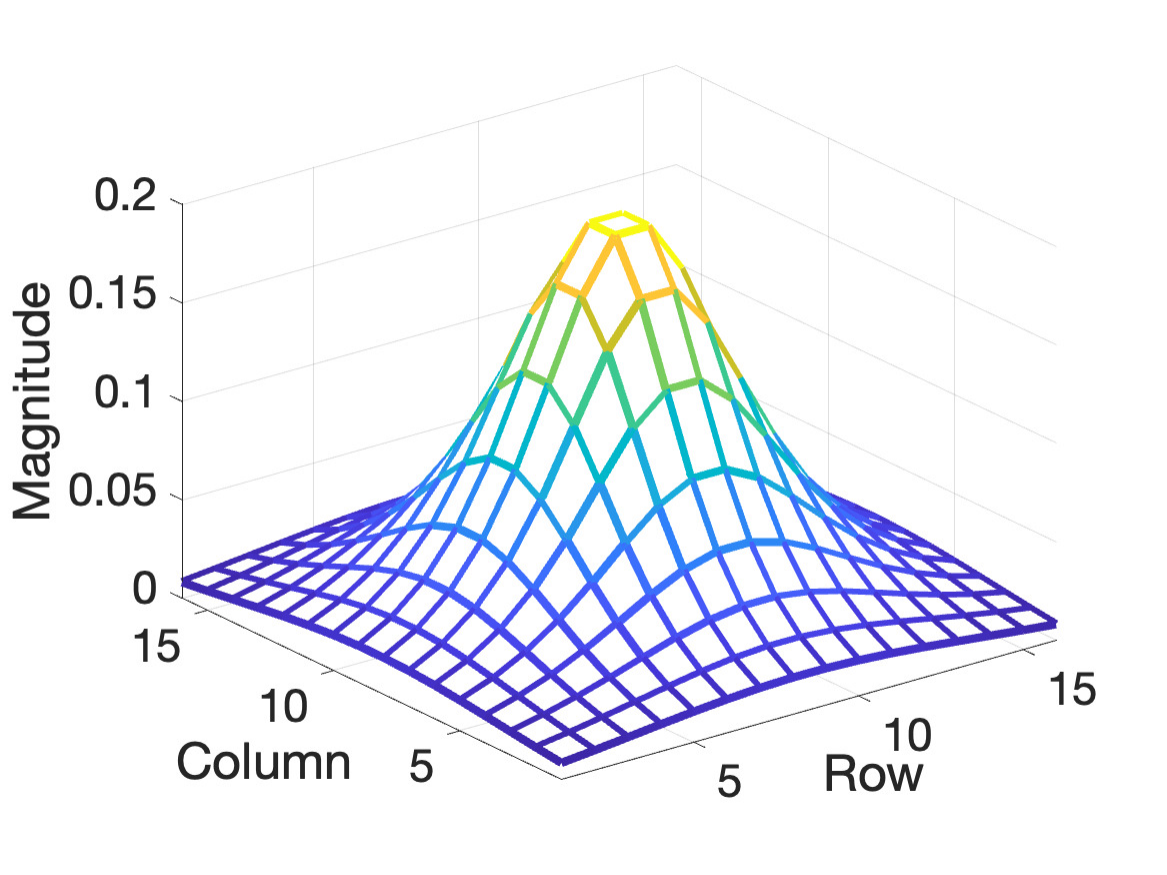}}
\caption{RIS PEM amplitude profile for a $16\times 16$ RIS with $2\times 2$ AMAF at distance $F = 6$.}
\label{fig:RIS_Ex2x2}
\vspace{-0.5cm}
\end{figure}

Consider the Singular Value Decomposition (SVD) of $\Tm=\Um\Sm\Vm^\herm$ where $\Um \in \CC^{N_p \times N_p}$ and $\Vm \in \CC^{N_a \times N_a}$ are unitary matrices and $\Sm \in \CC^{N_p \times N_a}$ is a diagonal matrix containing ordered singular values $\sigma_1, \sigma_2, \twodots, \sigma_{N_a}$. Letting $\uv_\ell, \vv_\ell$ denotes the $\ell^{\text{th}}$ columns $\Um, \Vm$, respectively, any AMAF weight vector $\bv \in \CC^{N_a}$ can be written as
$\bv = \sum_{\ell=1}^{N_a} \beta_\ell \vv_\ell$ with the transmit power
normalization $\sum_{\ell=1}^{N_a}  |\beta_\ell|^2 = 1$.
This results in the complex amplitude profile $\uv = \Tm \bv = \sum_{\ell=1}^{N_a} \sigma_\ell \beta_\ell \uv_\ell$. As an effective pragmatic (maximum power transfer) choice  of the AMAF weight vector, 
we let $\bv = \vv_1$, referred to as {\em principal eigenmode} (PEM) design. 
This results in the ``template'' RIS weight vector
$\tilde{\wv}_0 = \sigma_1 \uv_1 \odot e^{-j \angle \uv_1}$ with all elements in $\RR_+$, corresponding to a beam pointing in the boresight direction of the RIS.\footnote{Here $\angle \uv_1$ is the vector of phases of $\uv_1$, $\exp$ is applied componentwise, 
and $\odot$ is elementwise product.}
For example, Fig. \ref{fig:CenterSpot0} shows the template beam ground footprint for a 16x16 RIS center fed by a 2x2 AMAF at a 20m high base station mast with $\alpha=37.37\degree$ as detailed in Section \ref{sec:designs}. The PEM design yields the RIS amplitude taper (13.92 dB) shown in Fig.~\ref{fig:RIS_Ex2x2}, yielding -35 dB sidelobes and 27.5 dBi gain. 
Beamsteering of the template beam to a desired direction $(\phi_0,\theta_0)$ 
is obtained by imposing a linear phase gradient in the form ${\wv} = \av(\phi_0,\theta_0) \odot \tilde{\wv}_0$. 

\begin{rem}
It is important to notice that the AMAF weight vector depends only on the AMAF-RIS geometry of the basic module, and not on the beam-steering. In a multi-beam multiuser setting, only the steering depends on the user channels and can be adapted by changing the 
RIS phases while the AMAF configuration remains fixed. \hfill $\lozenge$ 
\end{rem}

\section{Multi-beam multiuser MIMO}  
\label{subsec:HybridBF}

Following the``One Stream Per Subarray'' (OSPS) approach to 
multiuser MIMO \cite{commit_bf1}, we consider stacking 
$K > 1$ AMAF-RIS basic modules with minimal separation of length $\lambda/2$ (i.e., unit separation in normalized distance) for maximum space utilization. 

The global matrix $\mathbb{T} \in \CC^{K N_p \times K N_a}$ for the stacked modules can be written 
as a $K \times K$ block matrix where each block $\Tm_{i,j} \in \CC^{N_p \times N_a}$ models the propagation between the $j$-th AMAF and the $i$-th RIS and is calculated from \eqref{Telement}. We define also
$\Wm_i = \diag(\wv_i)$, for the $i$-th RIS phase-only weight vector (not inclusive of the 
amplitude taper as in $\tilde{\wv}$) and $\bv_j \in \CC^{N_a \times 1}$ to be the $j$-th AMAF weight vector. Defining the block-diagonal matrices  $\Wm = \diag(\Wm_i : i = 1, \ldots, K)$ and $\Bm = \diag(\bv_j : j = 1, \ldots, K)$, the global 
transmission matrix $\Nm \in \CC^{K N_p \times K}$ from the $K$ baseband antenna ports (each driving one AMAF) and the RIS reflecting elements is given by $\Nm = \Wm \mathbb{T} \Bm$ with $(i,j)$-th $N_p \times 1$ blocks
%\eqref{eq:DTB2}, where the off-diagonal blocks capture NEXT effects.
%\vspace{-12pt}
\begin{equation}  \label{eq:DTB2}
\left [ \Nm \right ]_{i,j} =\Wm_i \Tm_{i,j} \bv_j. 
\end{equation}
Since  the $K$ modules are identical, we have that 
$\Tm_{1,1} = \cdots = \Tm_{K,K}$. Our pragmatic design uses the PEM approach for each module, i.e.,$\bv_j = \vv_1$ for all $j = 1,\ldots, K$. Hence, the diagonal blocks $\Nm_{j,j}$ (in isolation) produce radiation patterns as seen before, 
i.e., independently steered versions of the basic weight vector $\tilde{\wv}_0$.
The off-diagonal blocks $\Nm_{i,j}$ for $i \neq j$ capture the effect of the near-end crosstalk (NEXT) between the modules. 
Fortunately, given the good tapered profile of the PEM design (see Fig.~\ref{fig:RIS_Ex2x2}), it turns out that the NEXT is essentially 
negligible even for unit separation between the modules. 

We consider a multiuser MIMO scenario where the base station (BS) is 
equipped with $K$ AMAF-RIS modules and serves multiple users located in the coverage area (see Fig.~\ref{3D-fig11}).  
The downlink scheduler chooses groups of $K$ users to be served on the same time slot by spatial multiplexing. The resulting LOS baseband channel matrix $\Hm \in  \CC^{K \times K}$ between the $K$ BS antenna ports and the $K$ (far-field) users is given by 
\begin{equation}
\label{eq:BB_MIMO}
    \Hm = \Am^\herm \Nm  = \Am^\herm \Wm \mathbb{T} \Bm ,   
\end{equation}
where 
\begin{equation}
\label{eq:Amat}
 \Am = 2 \left[\text{cos} \phi_1 \text{cos} \theta_1 \av(\phi_1,\theta_1), \ldots, \text{cos} \phi_K \text{cos} \theta_K \av(\phi_K,\theta_K) \right ] 
\end{equation}
is the $K N_p \times K$ matrix containing the steering vectors whose elements are given by (\ref{eq:array-vec}) from the overall stacked RIS array to the $K$ 
users, where each user $k$ is seen at an angle $(\phi_k,\theta_k)$ with respect to the 
RIS S2 coordinate system. Each $k$-th steering vector is weighted by 
$\sqrt{G_{\rm patch}(\phi_k,\theta_k)} = 2 \cos \phi_k \cos \theta_k$ due to the  
RIS element directivity. The off-diagonal terms in $\Hm$ capture the far-end crosstalk (FEXT) 
due to the sidelobes of the AMAF-RIS beams. 
Ideally, we want the matrix $\Hm$ to be strongly diagonal-dominant which allows us to dispense 
with (digital) baseband signal processing techniques such as zero-forcing. 
This can be achieved by a) our proposed PEM design which yields very low sidelobes, and 2) by scheduling sets of $K$ ``compatible'' users, which in the LOS MU-MIMO case means users
with sufficient angular separation in azimuth and/or elevation. 
Notice that the selection of compatible (nearly mutually orthogonal) 
sets of users in MU-MIMO is a common practice, as currently implemented in 802.11ax MU-MIMO mode (e.g., see \cite{wang2018scheduling} and references therein).

The achievable communication rate of user $k$ 
under Gaussian single-user capacity achieving codebooks and treating multiuser interference as noise is given by  
\begin{equation}
\label{eq:Rate}
 R_k = \log_2 \left(1+ \frac{| H_{k,k}|^2 P_{\rm RF}}{W N_0/L_k + \sum_{j=1, j \neq k}^{K}| H_{k,j}|^2 P_{\rm RF}}\right),
\end{equation}
bits per complex signal dimension, 
where $H_{k,j}$ is the $(k,j)$-th element of $\Hm$, $P_{\rm RF}$ is the total AMAF output RF power,  $N_0$ is the complex baseband AWGN power spectral density, $W$ is the channel bandwidth, and $L_k = (\lambda / (4\pi \rho_k))^2$  is the free-space pathloss due to distance $\rho_k$ (in meters) 
between user $k$ and the BS.  
In the LOS condition, $\Hm$ is deterministic, and hence any standard synchronization
(carrier frequency, timing, and phase) at the user receivers can easily achieve (almost) ideal coherent detection. 
%All the usual problems related to ``imperfect channel state information'' that arise in typical %wireless communication scenarios do not play a significant role here.  
%As we shall see in the performance example of Section \ref{sec:designs}, the matrices $\Hm$ 
%obtained with PEM precoding are strongly diagonally dominant. Since the multiuser interference %is essentially negligible, it does not make sense to attenuate the power of some data stream to %mitigate interference to other streams. 
%Indeed, for the example of this paper, the best choice consists of transmitting at maximum RF %output power from all the AMAFs. 

%%%%%%%%%%%%%%%%%%%%%%%%%%%%%%%%%%%%%%%%%%%%%%%%
\vspace{-5pt}
\section{MU-MIMO Example} 
\label{sec:designs}

We present a case study where the RIS and the AMAF are SRAs of size $N_x = N_z = 16$ and $N_h = N_v = 2$, respectively.  We choose an empirically optimum focal length $F = 6$, chosen to strike a good balance between beam directivity, loss in the AMAF-RIS structure (captured by the value of singular-value $\sigma_1$), and height of the sidelobes. 
Qualitatively, if $F$ is too small only a central portion of the RIS is illuminated by the AMAF, i.e., the elements of the RIS away from the center play no role in beamforming. If $F$ is too large, then a large fraction of the RF power radiated by the AMAF is lost in space (and creates significant NEXT in the multi-module stacked array), 
since the solid angle covered by the RIS is too small and also higher FEXT due to a smaller RIS taper. In the case at hand, $F = 6$ with the PEM beamforming yields the nice tapered amplitude profile shown in Fig.~\ref{fig:RIS_Ex2x2}, resulting in the template beam with footprint 
in Fig.~\ref{fig:CenterSpot0}. 

We consider $K = 4$ stacked modules height $h = 20$m 
on the ground, serving a sector on the ground S1 x-y plane with range between $r_{\rm min} = 10$m to $r_{\rm max} = 100$m,  azimuth $\phi$ from $-60\degree$ to $60\degree$. 
The 10m and 100m ground distances correspond to the downlook angles of $\alpha_{\rm max} = \text{acot}~ (r_{\rm min}/h) = 63.43\degree$ and $\alpha_{\rm min} = \text{acot}~ (r_{\rm max}/h) = 11.30\degree$, respectively, with respect to the S2 origin. 
Therefore, we choose the RIS mechanical downtilt angle $\alpha$ to be the arithmetic mean, i.e. $\alpha = 37.37 \degree$, for the optimum element factor utilization in the elevation.
This downtilt angle causes the RIS normal vector to intercept the ground at distance $r = 26.19$ m, as shown in Fig. \ref{fig:CenterSpot0}.

In order to ensure a minimum 0 dB\footnote{Notice that an SNR of 0 dB corresponds to a 
channel capacity equal to 1 per complex dimension, which is approachable in practice using QPSK modulation with powerful binary LDPC coding of rate (slightly less than) 1/2.
Hence, such a system is quite realistic also from a practical viewpoint.} 
SNR to any user in the picocell, we consider the link budget of Table \ref{tab:SRS}. The RF feed power from the AMAF, $P_{\rm RF}=P_T/G(\phi,\theta)$
%\footnote{This implies that $\sigma_1 = 1$. In the numerical simulations, we get $\sigma_1=1.08$. A $\sigma_1>1$ is not practically possible (in hardware) because AMAF-RIS propagation is a passive mechanism. Therefore, we use $\sigma_1 = 1$ in this numerical study. Full-wave simulation and hardware-based determination of $\sigma_1$ is an extension work.}, 
where $G(\phi,\theta)$ is given in \eqref{eq:bfpattern}. At the cell edge, $G(\phi=60\degree,\theta=26.06\degree)=20.7~{\rm dBi}$, yielding  
a required AMAF RF power of $P_{\rm RF}=40.7~{\rm dBm}-20.7~{\rm dBi}=20~{\rm dBm}$.

%\vspace{-8pt}
\begin{table}[h!]
\caption{Example system specifications.}
\label{tab:SRS}
\centering
\setlength{\tabcolsep}{2pt} 
\begin{tabular}{lclc}
\hline\hline %inserts double horizontal lines\\
Specification & Value & Specification & Value \\
\hline % inserts single horizontal line
Carrier freq. (GHz)       & 100 & Receive noise pow. (dBm)     & -72   \\
Cell range (m)            & 10 to 100  & Receive SNR (dB)  & 0 \\
Azimuth span ($\phi$)     & +/-60$\degree$ & Receive signal power (dBm) & -72 \\
Bandwidth $W$ (GHz) & 5  &  Path Loss $L_{\rm max}$ (dB) & 112.7  \\
Thermal noise pow. (dBm)  & -77  & EIRP $P_T$ (dBm)    & 40.7  \\
Rx NF (dB)      & 5  & RIS size ($N_x\times N_z$)  & 16 x 16  \\
\hline %inserts single line
\end{tabular}
\end{table}

\begin{figure}[h] 
\vspace{-0.3cm}
\centerline{\includegraphics[width=8cm,height=4cm]{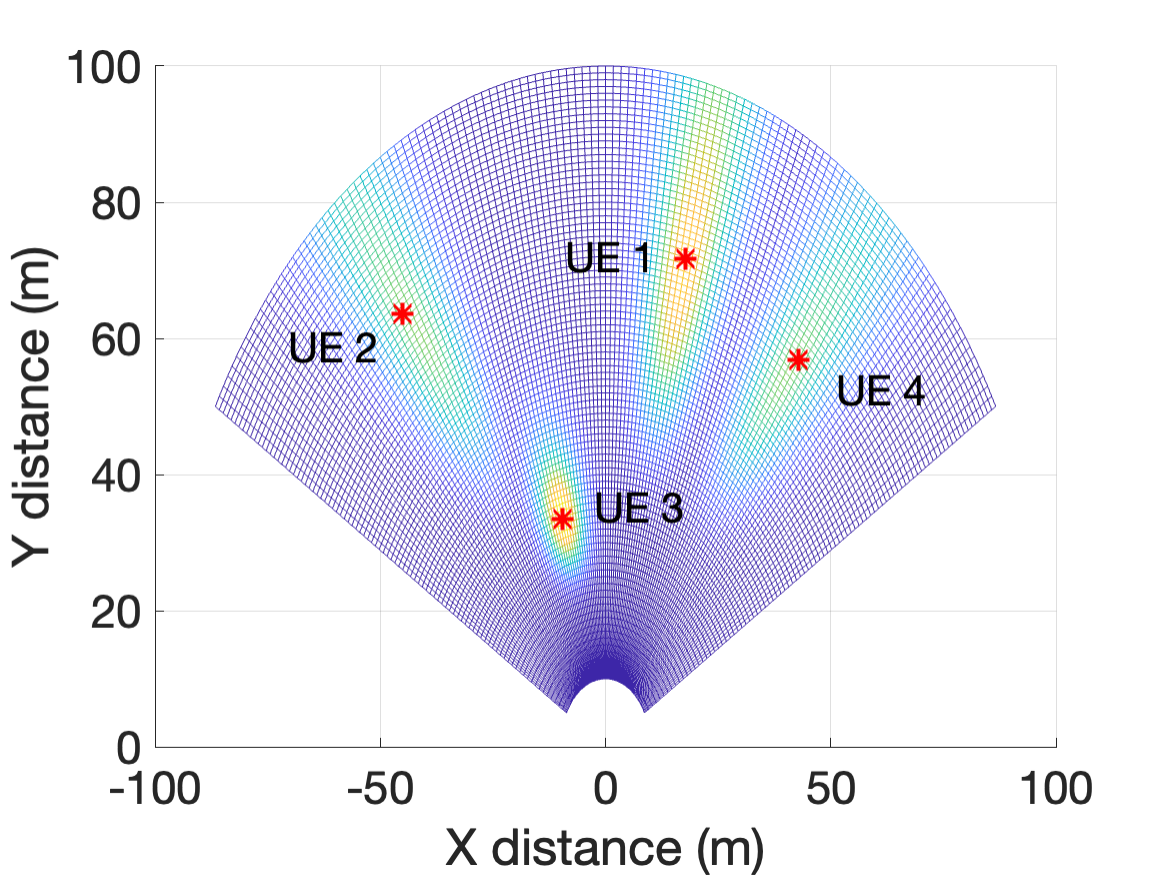}}
\caption{Ground footprints: an example set of 4 PEM spot beams, with perfect beam pointing onto the 4 UEs.}
\label{fig:MUspots}
\vspace{-0.5cm}
\end{figure}

We consider $K = 4$ downlink data streams serving 4 users randomly distributed with 
azimuth $\phi \in [-60\degree, 60\degree]$ and range $r \in [10\text{m}, 100\text{m}]$. 
The scheduler chooses the $K$ users a minimum azimuth 
angle separation of $15\degree$, corresponding to the -20 dB beam contour of the template beam. 
Fig. \ref{fig:MUspots} shows a snapshot (random realization) of 4 user positions and the corresponding ground beam footprints with ideal beam steering (i.e., by pointing the beams towards the corresponding users angles).

%\vspace{-10pt}
\begin{figure}[h] 
\vspace{-0.2cm}
\centerline{\includegraphics[width=8cm,height=4cm]{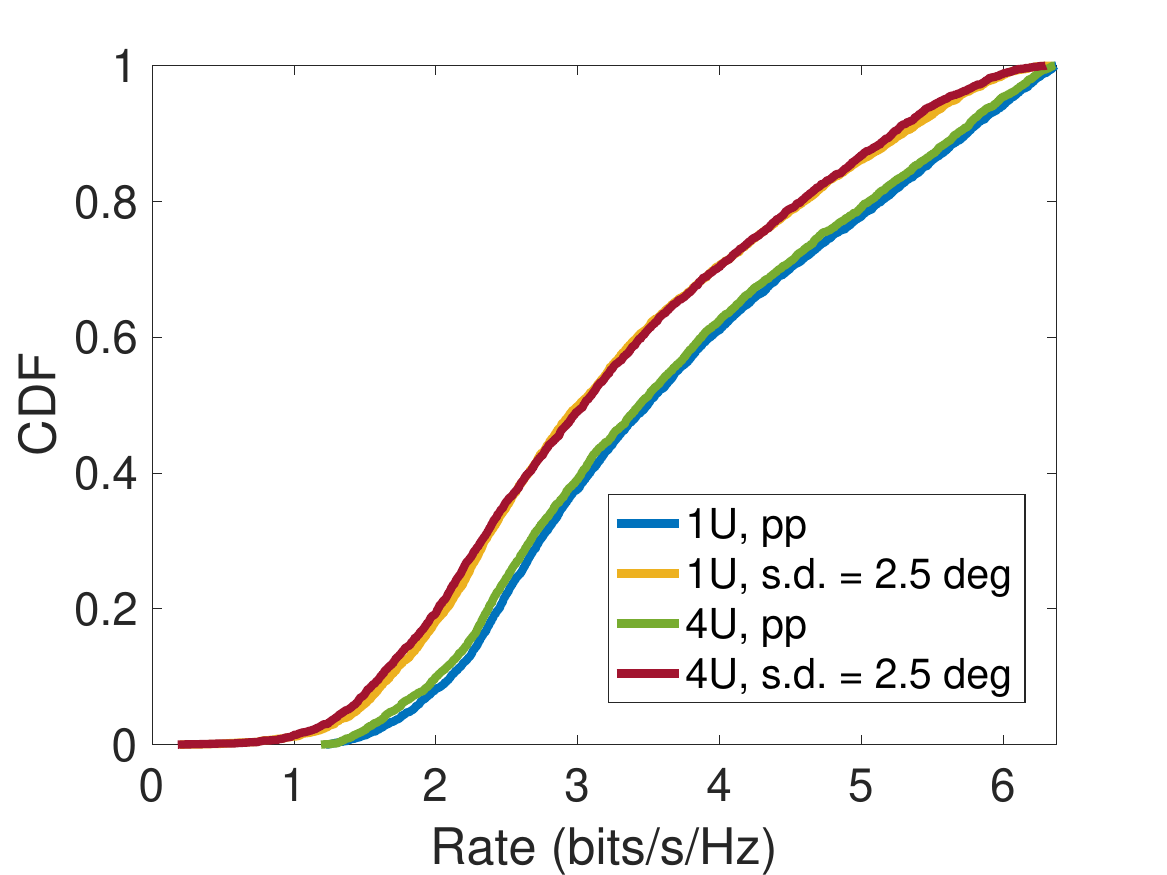}}
\caption{Rate CDFs for the beam pointing case, with perfect pointing (pp) and with Gaussian beam pointing errors.}
\label{fig:PP_rate_CDF}
\vspace{-0.2cm}
\end{figure}

Fig. \ref{fig:PP_rate_CDF} shows the achievable rate CDF for the case of a single user (1U),  and 4 users (4U), with perfect beam pointing 
(pp), and independent Gaussian beam pointing errors in both $\phi$ and $\theta$ 
with standard deviations of $2.5 \degree$. We find that the rate degradation due to beam pointing errors is tolerable because the beam footprints have a smooth contour, which provides some robustness to pointing errors. We see also that there is no practical degradation between the single user and the 4 users case, indicating that the FEXT/NEXT multiuser interference is effectively negligible. 
This means that any further hybrid precoding would yield no improvement at the cost of
a much higher computational complexity in the baseband. 

\begin{figure}[h]
\vspace{-0.5cm}
\centerline{\includegraphics[width=8cm,height=4cm]{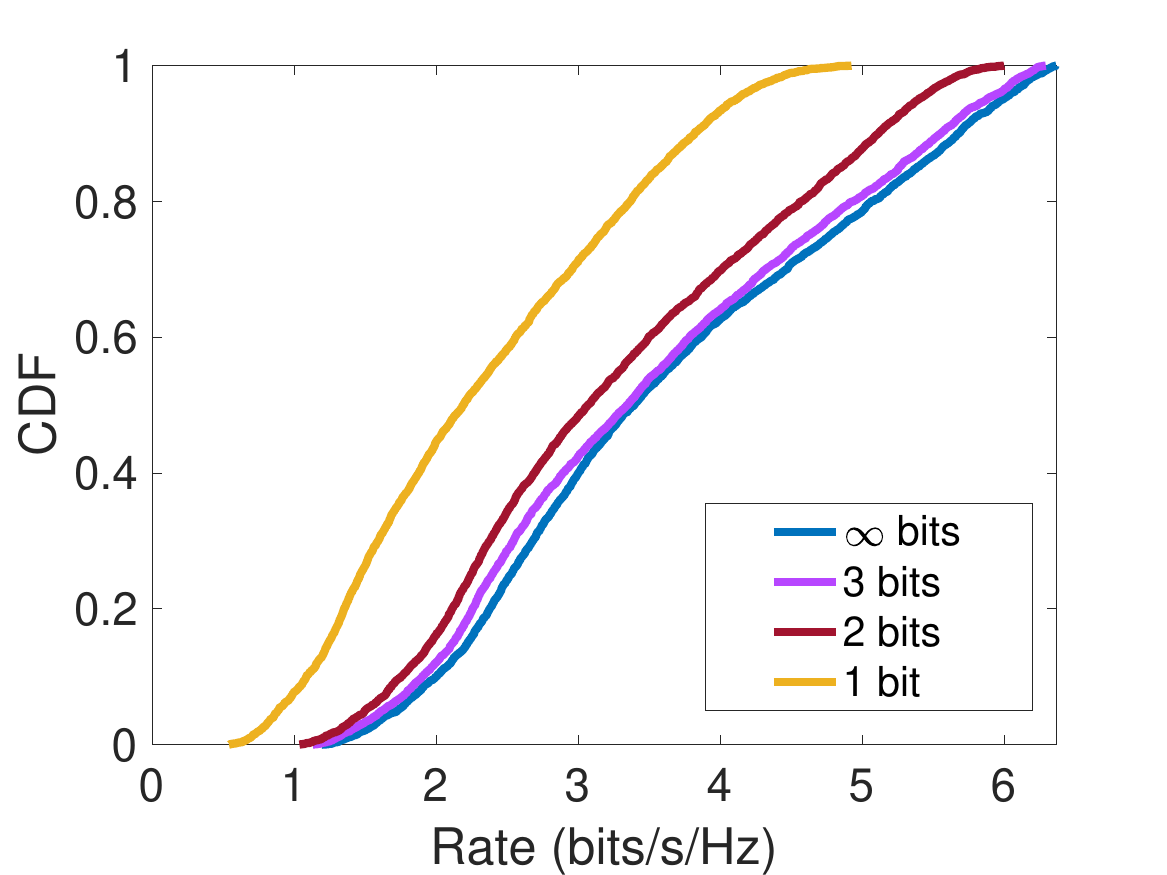}}
\caption{Rate CDFs for the beam pointing case, with perfect pointing and RIS quantized phase-shifters.}
\label{fig:Quant_MU}
\vspace{-0.3cm}
\end{figure}

In Fig. \ref{fig:Quant_MU} we show the rate CDF for the same system scenario, with no pointing errors and quantized RIS phase-shifters. We used simple scalar quantization of the continuous phases, without any further complicated optimization. 
For 3 or more quantization bits, the rate CDF is essentially identical to the unquantized case. Nevertheless, even for just 1 bit quantization, i.e., restricting the RIS phase shifts to $\pm 1$, the per-user rates range from $\approx 0.5$ to $\approx 5$ bits/symbol, corresponding to 
$\approx 2.5$ to $\approx 25$ Gb/s (with $W = 5$ GHz as in Table \ref{tab:SRS}).

%%%%%%%%%%%%%%%%%%%%%%%%%%%%%%%%%%%%%%%%%
\vspace{-5pt}
%%%%%%%%%%%%%%%%%%%%%%%%%%%%%%%%%%%%%%%%%%%%%%%%
\section{Power Efficiency Analysis} 

\begin{figure}[h] 
\vspace{-0.47cm}
\centerline{\includegraphics[width=3.355cm]{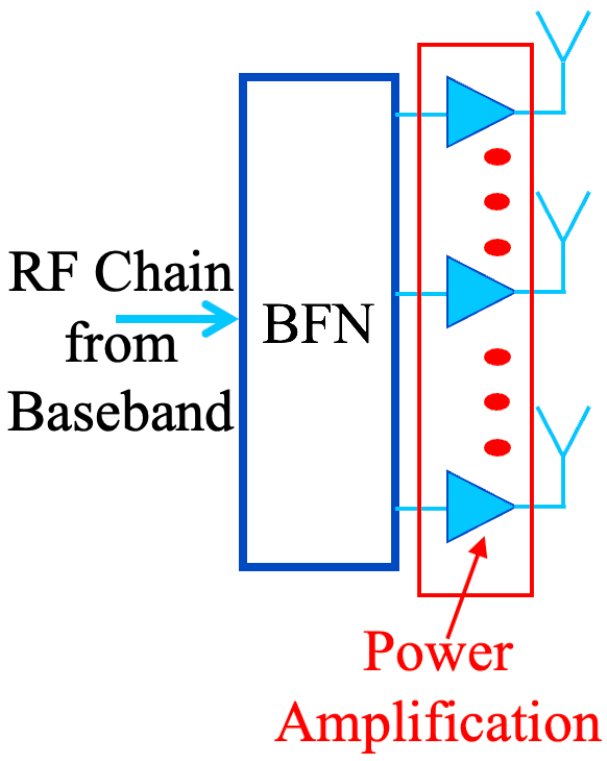}}
\caption{Arch.2 block diagram. Beamforming network (BFN) consists of amplitude and phase shifters.}
\label{fig:Arch.2}
\vspace{-0.15cm}
\end{figure}

For the same planar array dimension and 
beamforming radiation pattern of the proposed AMAF-RIS architecture (Arch.1), we consider 
a baseline architecture consisting of a constrained-fed active array (Arch.2), where each element has its own dedicated power amplifier (PA), see Fig. \ref{fig:Arch.2}.
For Arch.2 we neglect the power loss incurred by the beamforming network from the antenna port to the array elements (i.e., to the signals before the per-antenna PAs)
because this operates on low-level signals and impacts essentially only the noise figure, which we assume here to be ideal. This assumption is favorable to Arch.2. Nevertheless, we shall see that Arch.1 (the proposed one) is still very competitive from the energy efficiency viewpoint. 

From the previously computed link budget in Section \ref{sec:designs}, we target 0 dB SNR for a cell edge user, given by ${\rm SNR}_{\rm rx} = P_{\rm RF}G(\phi=60\degree,\theta=26.06\degree)/(L_{\rm max}WN_0)$. Plugging in the values from Section \ref{sec:designs}, we get $P_{\rm RF}=20~{\rm dBm}$. For Arch.1, the AMAF weights, $\vv_1=[0.5, 0.5, 0.5,0.5]$. Hence, the maximum AMAF PA output power  
is given by $P^{(1)}_{\rm pa-max}={\rm max}|v_{1i}|^2P_{\rm RF}=-6~\text{dB}+20~\text{dBm}=14~\text{dBm}=25.1~\text{mW}$. We assume that all the PAs in the (AMAF) array are developed in the same semiconductor technology, and are all biased with the same DC power dictated by the maximum requested RF power. Considering Indium Phosphide (InP) PAs with efficiency $\eta=0.3$ \cite{ETH_PA_Survey, PAlimits},  the Arch.1 DC power consumption, $P^{(1)}_{\rm DC} = N_a P^{(1)}_{\rm pa-max}/\eta = 4 \times 25.1~\text{mW}/0.3 = 0.33~ \text{W}$. Likewise, for Arch.2, $P^{(2)}_{\rm {pa-max}}={\rm max}|u_{1i}|^2 P_{\rm RF}=-14.65~\text{dB}+20~\text{dBm}=5.34~\text{dBm}=3.42~\text{mW}$. Thus, Arch.2 DC power $P^{(2)}_{\rm DC} = N_p P^{(2)}_{\rm pa-max}/\eta=2.92~\text{W}$.
We see that the proposed architecture is almost 10x times more power efficient than the baseline. 

%%%%%%%%%%%%%%%%%%%%%%%%%%%%%%%%%%%%%%%%%%%%%%%%
\vspace{-8pt}
\section{Conclusions}  
\label{sec:CONC}
\vspace{-1pt}
We proposed a novel multiuser multibeam architecture with 
over-the-air active array-based feeding (AMAF) and RIS-based beam steering suited to very high frequency bands. The scheme is based on a fundamental ``module'' formed by an AMAF-RIS pair, with fixed geometry, and can accommodate any suitable number $K$ of independently steered data streams by stacking such module in a larger array. 
We demonstrated a design example with ($2\times 2$) active antennas at the AMAF and 
$16 \times 16$ passive elements at the RIS. We also demonstrated that there is no dramatic performance loss with practical hardware constraints such as quantized RIS phase shifters (3 bits or more) and beam pointing errors. Our pragmatic design approach is very simple, does not require complicated on-line optimization (unlike most analog-digital multiuser precoding approaches), and can be easily applied to different combinations of 
AMAF and RIS. The proposed architecture has low hardware complexity (very small number of PAs, simple active beamforming network), and achieves large energy efficiency gains with respect to the baseline active array design with the same beamforming capability. 

\small 
%%%%%%%%%%%%%%%%%%%%%%%%%%%%%%%%%%%%%%%%%%%%%%%%
\vspace{-2pt}
\section*{Acknowledgment}
\vspace{-2pt}
The work of G. Caire was supported by BMBF Germany in the program of ``Souverän. Digital. Vernetzt.'' Joint Project 6G-RIC (Project IDs 16KISK030).

%%%%%%%%%%%%%%%%%%%%%%%%%%%%%%%%%%%%%%%%%%%%%%%%
%\vspace{-5pt}
\bibliographystyle{IEEEtran}
%\vspace{-6pt}
\bibliography{P2-bibliography}
\vspace{-5pt}
\end{document}

%% file: macros.tex
\newcommand{\twodots}{\mathinner {\ldotp \ldotp}}
\newfont{\bbb}{msbm10 scaled 700}

\newfont{\bb}{msbm10 scaled 1100}
\newcommand{\CC}{\mbox{\bb C}}

\newcommand{\RR}{\mbox{\bb R}}

\newcommand{\av}{{\bf a}}
\newcommand{\bv}{{\bf b}}

\newcommand{\iv}{{\bf i}}
\newcommand{\jv}{{\bf j}}
\newcommand{\kv}{{\bf k}}

\newcommand{\nv}{{\bf n}}

\newcommand{\pv}{{\bf p}}

\newcommand{\uv}{{\bf u}}
\newcommand{\wv}{{\bf w}}
\newcommand{\vv}{{\bf v}}

\newcommand{\Am}{{\bf A}}
\newcommand{\Bm}{{\bf B}}

\newcommand{\Hm}{{\bf H}}

\newcommand{\Nm}{{\bf N}}

\newcommand{\Sm}{{\bf S}}
\newcommand{\Tm}{{\bf T}}
\newcommand{\Um}{{\bf U}}
\newcommand{\Wm}{{\bf W}}
\newcommand{\Vm}{{\bf V}}

\newcommand{\diag}{{\hbox{diag}}}

\newcommand{\herm}{{\sf H}}

\newcommand{\transp}{{\sf T}}

\usepackage{stmaryrd} % for \mkv 

\usepackage{hyperref}
\hypersetup{
    bookmarks=true,         % show bookmarks bar?
    unicode=false,          % non-Latin characters in AcrobatÕs bookmarks
    pdftoolbar=true,        % show AcrobatÕs toolbar?
    pdfmenubar=true,        % show AcrobatÕs menu?
    pdffitwindow=false,     % window fit to page when opened
    pdfstartview={FitH},    % fits the width of the page to the window
    pdfnewwindow=true,      % links in new window
    colorlinks=true,       % false: boxed links; true: colored links
    linkcolor=red,          % color of internal links (change box color with linkbordercolor)
    citecolor=green,        % color of links to bibliography
    filecolor=blue,      % color of file links
    urlcolor=blue           % color of external links
}